# Fokker-Planck formalism in magnetic resonance simulations

Ilya Kuprov

School of Chemistry, University of Southampton, University Road, Southampton, SO17 1BJ, UK.

i.kuprov@soton.ac.uk

## Abstract

This paper presents an overview of the Fokker-Planck formalism for non-biological magnetic resonance simulations, describes its existing applications and proposes some novel ones. The most attractive feature of Fokker-Planck theory compared to the commonly used Liouville - von Neumann equation is that, for all relevant types of spatial dynamics (spinning, diffusion, flow, *etc.*), the corresponding Fokker-Planck Hamiltonian is time-independent. Many difficult NMR, EPR and MRI simulation problems (multiple rotation NMR, ultrafast NMR, gradient-based zero-quantum filters, diffusion and flow NMR, off-resonance soft microwave pulses in EPR, spin-spin coupling effects in MRI, *etc.*) are simplified significantly in Fokker-Planck space. The paper also summarises the author's experiences with writing and using the corresponding modules of the *Spinach* library – the methods described below have enabled a large variety of simulations previously considered too complicated for routine practical use.

# 1. Introduction

Good theory papers have two essential features: they are readable and computable. That is the reason why the Liouville - von Neumann equation (on the coherent side) and Bloch-Redfield-Wangsness / Lipari-Szabo theories (on the relaxation side) dominate magnetic resonance – there are papers and books that describe them with the eloquence and elegance of a well written detective story [1-4].

In this respect, Fokker-Planck theory is a hard sell. Its applications to magnetic resonance feature impractically large matrices [5], highly technical derivations [6], and very flexible dynamical models that require a level of knowledge few people would ever have about their systems [7]. Yet the prize is tempting – Fokker-Planck theory captures, in one general equation, almost everything there is to model in NMR, EPR and MRI. Most importantly, it treats spatial dynamics at the same conceptual level as spin dynamics, and includes relaxation processes in a natural way that is free of perturbative assumptions [8].

The problem with the *status quo* is that spatial degrees of freedom are too often an afterthought in the Liouville - von Neumann formalism. For condensed phase spin systems it is commonly assumed that the spin state has no influence on spatial dynamics, but that spatial dynamics has an effect on the spin Hamiltonian [1,2]. In other words, it is assumed that "something happens" to spatial coordinates that makes the spin Hamiltonian time-dependent. The resulting equation of motion is

$$\frac{\partial}{\partial t}\boldsymbol{\rho}(t) = -i\mathbf{H}(t)\boldsymbol{\rho}(t) \tag{1}$$

where $\boldsymbol{\rho}(t)$ is the density operator and $\mathbf{H}(t)$ is the spin Hamiltonian commutation superoperator. When the Hamiltonian contains stochastic terms (for example, from rotational diffusion in liquids), the effect of those terms is represented by a relaxation superoperator $\mathbf{R}$ [1,3,9] that can be modified to drive the solution to some thermal equilibrium state [10,11]:

$$\frac{\partial}{\partial t}\boldsymbol{\rho}(t) = \left[-i\bar{\mathbf{H}}(t) + \mathbf{R} + \mathbf{K}\right]\boldsymbol{\rho}(t) \tag{2}$$

where the overbar indicates ensemble averaging and $\mathbf{K}$ is the kinetics superoperator that accounts for the possible presence of chemical processes in the system [2]. This equation is currently the central pillar of most magnetic resonance simulation frameworks [12-17]. It is deterministic, and many methods exist (Floquet theory [18], time slicing [15], *COMPUTE* [19], *etc.*) for calculating its exact solution analytically or numerically. The biggest source of complications here is that the "invisible hand" of spatial dynamics makes even the ensemble-averaged spin Hamiltonian time-dependent in non-trivial ways. This can lead to spectacularly complex analytical solutions – the excellent Equation 38 in the recent paper by Scholz, Meier and Ernst [20] is a good indication that perhaps the time has come to take a closer look at a numerical formalism that promises to avoid it.

The most attractive feature of Fokker-Planck theory compared to the Liouville - von Neumann formalism is that, *for all common types of spatial dynamics (spinning, diffusion, flow, etc.), the corresponding Fokker-Planck evolution generator is time-independent*. It is demonstrated below that many difficult magnetic resonance simulation problems (multiple rotation NMR, gradient chirps, ultrafast NMR, soft off-resonance microwave pulses in EPR, overtone NMR, *etc.*) are simplified significantly in Fokker-Planck space. It also produces major generalisations and simplifications across the simulation code – recent versions of *Spinach* [12] owe their versatility to the Fokker-Planck formalism.

## 2. Fokker-Planck formalism in spin dynamics

### 2.1 Fokker-Planck equation

The state of an individual system in classical physics can be described by a vector of state variables that we will denote $\mathbf{x}$. The associated equation of motion involves time derivatives and other operators acting on this vector. This formalism goes back to Newton [21], but describing an ensemble of systems in this way is technically difficult because the number of state variables becomes very large. A less detailed, but more convenient description of a classical ensemble may be formulated in terms of the probability density $p(\mathbf{x},t)$ of finding a system in a state $\mathbf{x}$ at time $t$. As the ensemble evolves, this probability flows around the state space in a way that depends on the interactions present. The resulting probability flux $\mathbf{j}(\mathbf{x},t)$ is determined by the local velocity $\mathbf{v}(\mathbf{x},t)$

$$\mathbf{j}(\mathbf{x},t) = \mathbf{v}(\mathbf{x},t) p(\mathbf{x},t) \tag{3}$$

and the velocity depends on the equation of motion.

The Fokker-Planck equation, proposed independently by Adriaan Fokker [22] and Max Planck [23], can be formulated as the continuity equation for the probability flux:

$$\frac{\partial p(\mathbf{x},t)}{\partial t} = -\nabla_{\mathbf{x}} \cdot \mathbf{j}(\mathbf{x},t) \tag{4}$$

It essentially means that probability cannot be destroyed or created. It can only be moved around: the local rate of decrease in the probability density is equal to the divergence of its flux.

### 2.2 Spin degrees of freedom

In situations where spin degrees of freedom are present in the system, Equation (4) may be extended to include the quantum mechanical density matrix $\boldsymbol{\rho}$ as a state variable:

$$\frac{\partial p(\mathbf{x},\boldsymbol{\rho},t)}{\partial t} = -\nabla_{\mathbf{x}} \cdot \left[\mathbf{v}_{\mathbf{x}}(\mathbf{x},\boldsymbol{\rho},t) p(\mathbf{x},\boldsymbol{\rho},t)\right] - \nabla_{\boldsymbol{\rho}} \cdot \left[\mathbf{v}_{\boldsymbol{\rho}}(\mathbf{x},\boldsymbol{\rho},t) p(\mathbf{x},\boldsymbol{\rho},t)\right] \tag{5}$$

The velocity in the spin space is given by Equation (1); the velocity in the lab space is determined by whatever is happening there. The average density matrix $\boldsymbol{\rho}(\mathbf{x},t)$ at every point $\mathbf{x}$ in the lab space is calculated by taking an integral over the probability density with respect to the spin degrees of freedom:

$$\boldsymbol{\rho}(\mathbf{x},t) = \int \boldsymbol{\rho} \, p(\mathbf{x},\boldsymbol{\rho},t) \, dV_{\boldsymbol{\rho}} \tag{6}$$

We need the equation of motion for this quantity; it is obtained by differentiating Equation (6) with respect to time, then using the expression for the derivative from Equation (5), and then going through a few rounds of simplifications. That is a surprisingly convoluted procedure – Appendix A in [5] contains a detailed discussion. Here we would simply present the final result:

$$\frac{\partial \boldsymbol{\rho}(\mathbf{x},t)}{\partial t} = -i\mathbf{L}(\mathbf{x},t)\boldsymbol{\rho}(\mathbf{x},t) + \mathbf{M}(\mathbf{x},t)\boldsymbol{\rho}(\mathbf{x},t) \tag{7}$$

in which $\mathbf{L}(\mathbf{x},t) = \mathbf{H}(\mathbf{x},t) + i\mathbf{R} + i\mathbf{K}$ is the Liouvillian that is responsible for spin dynamics and $\mathbf{M}(\mathbf{x},t)$ is the spatial dynamics generator that controls diffusion, flow, sample spinning, and other types of classical mechanics in the laboratory space. This is the equation of motion that will be referred to as the "Fokker-Planck equation" for the rest of this paper. In the special case when $\mathbf{M}(\mathbf{x},t)$ is the diffusion operator, Equation (7) is known as the stochastic Liouville equation [24].

## 2.3 Spatial dynamics generators

Expressions for $\mathbf{M}(\mathbf{x},t)$ come directly from classical mechanics [25]. A few relevant equations of motion for the probability density are (the spatial dynamics generator that we need is in square brackets):

1. Flow along a coordinate $x$ and circular motion with a phase $\varphi$

$$\frac{\partial p(x,t)}{\partial t} = \left[ v(t)\frac{\partial}{\partial x} \right] p(x,t), \qquad \frac{\partial p(\varphi,t)}{\partial t} = \left[ \omega(t)\frac{\partial}{\partial \varphi} \right] p(\varphi,t) \qquad (8)$$

   where $v$ is linear velocity and $\omega$ is angular velocity.

2. Circular motion around a specific axis $\mathbf{n} = \begin{bmatrix} n_X & n_Y & n_Z \end{bmatrix}$ in three dimensions

$$\frac{\partial p(\mathbf{r},t)}{\partial t} = \left[ \omega(t)\left(n_X \mathbf{J}_X + n_Y \mathbf{J}_Y + n_Z \mathbf{J}_Z\right) \right] p(\mathbf{r},t) \qquad (9)$$

   where $\{\mathbf{J}_X, \mathbf{J}_Y, \mathbf{J}_Z\}$ are angular momentum operators and $\omega$ is the angular velocity.

3. Translational diffusion in a potential $U(\mathbf{r})$ in an isotropic medium

$$\frac{\partial p(\mathbf{r},t)}{\partial t} = \left[ \frac{\sigma^2 \nabla^2}{2\gamma^2} - \frac{\nabla \cdot \mathbf{f}(\mathbf{r})}{\gamma} \right] p(\mathbf{r},t), \qquad \mathbf{f}(\mathbf{r}) = -\nabla U(\mathbf{r}) \qquad (10)$$

   where $\sigma$ is the amplitude of the fluctuating force that generates the diffusion and $\gamma$ is the friction constant of the medium.

4. Isotropic rotational diffusion in three dimensions

$$\frac{\partial p(\Omega,t)}{\partial t} = \left[ D_R \left( \mathbf{J}_X^2 + \mathbf{J}_Y^2 + \mathbf{J}_Z^2 \right) \right] p(\Omega,t) \qquad (11)$$

   where $\Omega$ is a shorthand for Euler angles or other orientation parameters, $\{\mathbf{J}_X, \mathbf{J}_Y, \mathbf{J}_Z\}$ are angular momentum operators and $D_R$ is the rotational diffusion coefficient.

Probability density evolution equations for other types of motion may be found in the specialist literature [26,27]. For our purposes here, these equations yield the spatial dynamics generator that is to be inserted into Equation (7). For example, the stochastic Liouville equation used in electron spin resonance [24] is obtained by using the rotational diffusion term:

$$\frac{\partial \boldsymbol{\rho}(\Omega,t)}{\partial t} = -i\mathbf{L}(\Omega,t)\boldsymbol{\rho}(\Omega,t) + D_R \left( \mathbf{J}_X^2 + \mathbf{J}_Y^2 + \mathbf{J}_Z^2 \right) \boldsymbol{\rho}(\Omega,t) \qquad (12)$$

Although spatial dynamics generators can depend on time, they are usually static in practice. One important thing to note is that this includes not only diffusion, but deterministic motion as well. For example, magic angle spinning and phase rotation of an external radiofrequency field both have time-independent generators within the Fokker-Planck formalism.

Quantum mechanical processes involving spatial degrees of freedom, such as quantum rotor [28] and quantum oscillator [29] dynamics, may be accounted for by using the Schrödinger Hamiltonian as the generator. In that case, the Fokker-Planck equation yields back the Liouville - von Neumann formalism.

## 2.4 Matrix representations of spatial dynamics generators

Analytical solutions to Equation (7) can be complicated and uninviting; that is the primary reason for the frosty reception that the Fokker-Planck formalism has seen so far in the magnetic resonance community.

In this brief overview we would argue that *numerical* solutions, where spatial coordinates are discretised on finite grids [30], are easier to understand, implement and use because all differential operators of the kind shown in the previous section become matrices.

Of course, selecting a basis of continuous functions would also lead to a matrix representation – for example, Wigner *D*-functions are a popular choice for rotational dynamics problems [6,8,24]. Finite grids are advocated here because they are easier to automate – there are standard libraries that generate differential operators on any given grid. For uniform grids with periodic boundary conditions, Fourier differentiation matrices [31] are recommended:

$$\left[\frac{\partial}{\partial \varphi}\right]_{nk} = \begin{cases} \frac{(-1)^{n+k}}{2}\cot\left[\frac{(n-k)\pi}{N}\right] & n \neq k \\ 0 & n = k \end{cases} \tag{13}$$

Operators for higher derivatives are obtained by taking powers of this matrix. For non-periodic coordinates or non-uniform grids, finite difference matrices are convenient – a general algorithm for building them has been published by Fornberg [32]. Implementations of Equation (13) and of the Fornberg method are available in *Spinach* [12]. Fourier differentiation matrices are an exact representation of the derivative operator on a given uniform grid – using them for rotational dynamics is important because magic angle spinning NMR simulations can go through thousands of rotor cycles. The number of discretisation points required for each grid is dictated by the accuracy required. Formal error bounds are available [31,32], but in practice the grid size is simply increased until convergence is achieved on the output.

## 2.5 Algebraic structure of the composite problem

From the algebraic perspective, the Fokker-Planck equation operates in a direct product of spatial and spin coordinates. For logistical reasons (to do with data layouts in *Matlab*), it is preferable to have spatial coordinates as the first term in the Kronecker product and spin coordinates as the second term. Thus, the general form of a purely spatial operator is $\mathbf{A} \otimes \mathbf{1}$, the general form of a pure spin operator is $\mathbf{1} \otimes \mathbf{B}$ and the form of an operator that correlates spatial and spin degrees of freedom (for example, a pulsed field gradient) is $\mathbf{A} \otimes \mathbf{B}$. The spin subspace has the usual direct product structure [1,2,15] and for the spatial coordinates it is reasonable to put all periodic coordinates ahead of all non-periodic ones. All of this results in the following product algebra for the evolution generators

$$\left[\mathfrak{so}_2(\mathbb{R})\right]^{\otimes m} \otimes \left[\mathfrak{gl}_k(\mathbb{R})\right] \otimes \left[\mathfrak{su}_n(\mathbb{R})\right] \tag{14}$$

where the first term in the direct product accounts for the (possibly multiple) radiofrequency irradiations and uniaxial sample rotations, all generated by $\mathfrak{so}_2(\mathbb{R})$ Lie algebras, the second term is responsible for translational and diffusional dynamics generated by the general linear algebra $\mathfrak{gl}_k(\mathbb{R})$ acting on a $k$-dimensional real space, and the last term is the usual special unitary algebra $\mathfrak{su}_n(\mathbb{R})$ generating the dynamics of an $n$-state quantum system [33]. In human language:

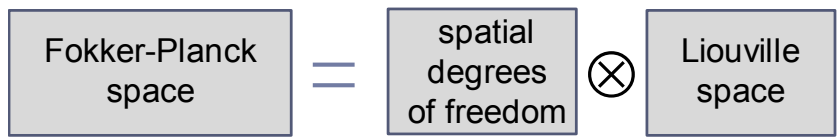

where the Liouville space [34-36] may be formally defined as the vectorisation of the density operator space acted upon by the adjoint Lie algebra of $\mathfrak{su}_n(\mathbb{R})$. Once the matrix representations for all operators are built, the Fokker-Plank equation of motion has the following general form:

$$\frac{\partial \boldsymbol{\rho}(\mathbf{x},t)}{\partial t} = -i\mathbf{F}(\mathbf{x},t)\boldsymbol{\rho}(\mathbf{x},t), \qquad \mathbf{F}(\mathbf{x},t) = \mathbf{L}(\mathbf{x},t) + i\mathbf{M}(\mathbf{x},t) \qquad (15)$$

The process of recovering the average spin density matrix from the Fokker-Planck state vector is particularly straightforward in *Matlab*: the matrix is reshaped to have spatial degrees of freedom along one dimension and spin degrees of freedom along the other, and summed over the spatial degrees of freedom.

The direct product structure in Equation (14) can produce matrices of large dimension. As an example, using ten points in the RF phase grid, 100 slices in the sample position grid and a five-spin system (1024 × 1024 matrices in Liouville space) would produce a Fokker-Planck matrix dimension of 10 × 100 × 1024 ≈ $10^6$. Because finite difference matrices and spin operators are very sparse (there are hard guarantees for spin Hamiltonians in the Pauli basis [37]), this is not a problem – as the examples given below demonstrate, modern state space reduction [38-41], sparse matrix manipulation [42] and diagonalization-free simulation [43,44] techniques are sufficient.

## 3. Case study 1: magic angle spinning NMR

Fokker-Planck theory is the primary working formalism for MAS NMR simulations in *Spinach* [12]. Two different formulations are possible [5]. In Cartesian coordinates in three dimensions, the spatial dynamics generator comes from Equation (9):

$$\frac{\partial}{\partial t}\boldsymbol{\rho}(\Omega,t) = \left[ -i\mathbf{L}(\Omega) + \omega_{\text{MAS}}\left(n_X \mathbf{J}_X + n_Y \mathbf{J}_Y + n_Z \mathbf{J}_Z\right) \right]\boldsymbol{\rho}(\Omega,t) \qquad (16)$$

$$\mathbf{L}(\Omega) = \mathbf{H}(\Omega) + i\mathbf{R} + i\mathbf{K}$$

where $\Omega$ are Euler angles or any other directional parameters, and $\mathbf{H}(\Omega)$ is the spin Hamiltonian commutation superoperator. To make the rotation generator $\omega_{\text{MAS}}(n_X \mathbf{J}_X + n_Y \mathbf{J}_Y + n_Z \mathbf{J}_Z)$ turn the spin system around the magic angle, the axis $\mathbf{n}$ must be directed appropriately, for example

$$\mathbf{n} = \begin{bmatrix} 1 & 1 & 1 \end{bmatrix}/\sqrt{3} \qquad (17)$$

Alternatively, we could observe that, once the spinning axis direction is chosen, the motion is actually uniparametric – the only spatial variable that changes with time is the rotor phase $\varphi$. This leads to a more elegant formulation with a simpler spatial dynamics generator coming from Equation (8):

$$\frac{\partial}{\partial t}\boldsymbol{\rho}(\varphi,t) = \left[ -i\mathbf{L}(\mathbf{n},\varphi) + \omega_{\text{MAS}}\frac{\partial}{\partial \varphi} \right]\boldsymbol{\rho}(\varphi,t) \qquad (18)$$

The physical meaning of $\partial/\partial \varphi$ is that of a phase increment generator – it is responsible for moving the rotor phase forward at the rate $\omega_{\text{MAS}}$:

$$\exp\left[\omega_{\text{MAS}}\frac{\partial}{\partial \varphi}t\right]f(\varphi) = f(\varphi + \omega_{\text{MAS}}t) \qquad (19)$$

Selecting a uniform periodic phase grid $\{\varphi_j\}$ makes this operator a constant matrix given by Equation (13) and results in the following expression for the spin Liouvillians at the phase grid points:

$$\mathbf{L}(\mathbf{n},\varphi_j,\alpha,\beta,\gamma) = \sum_{km}\left[\mathfrak{D}_{\text{lab2rot}}(\mathbf{n})\mathfrak{D}_{\text{rotor}}(\varphi_j)\mathfrak{D}_{\text{crystal}}(\alpha,\beta,\gamma)\right]_{km}\mathbf{Q}_{km} + i\mathbf{R} + i\mathbf{K} \qquad (20)$$

where the 25 irreducible spherical components $\mathbf{Q}_{km}$ of the Hamiltonian are defined in our earlier paper dealing with a large-scale implementation of Redfield theory [44]. The best practical way of obtaining them is to use the Hamiltonian generation function of *Spinach* [12] that returns them as a cell array. The three sets of Wigner *D*-matrices accomplish the following roles. $\mathfrak{D}_{\text{crystal}}(\alpha,\beta,\gamma)$ rotates the molecule from the reference frame, in which the interactions are defined, into each of the individual crystallite orientations specified by the spherical averaging grid. $\mathfrak{D}_{\text{rotor}}(\varphi_j)$ then rotates the result, around the Z axis, into the orientation that corresponds to the specified rotor phase. $\mathfrak{D}_{\text{lab2rot}}(\mathbf{n})$ then rotates the result into the orientation that corresponds to the rotor direction vector $\mathbf{n}$. Finally, the matrix representation for the rotor turning generator $\partial/\partial\varphi$ is given by Equation (13). When the system dynamics permits, $\gamma$ angle averaging may be performed efficiently by averaging over the rotor phase in the same way as happens in Floquet theory [45]. Only a two-angle spherical grid is then required for powder averaging.

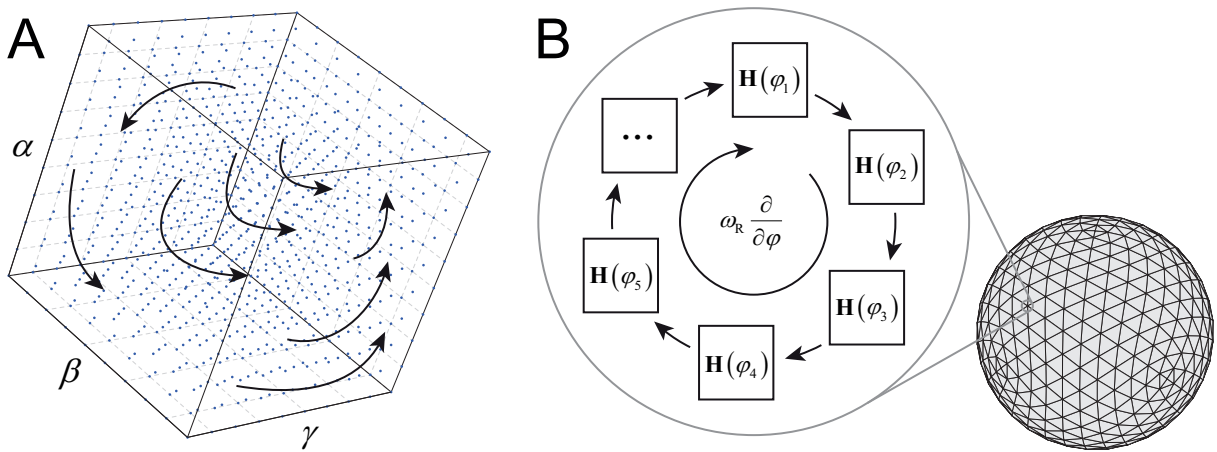

*Figure 1. A schematic illustration of the events taking place under (**A**) Equation (16), where the spatial dynamics operator generates a population flux through the entire orientation set and every point effectively hosts a Liouville - von Neumann equation for its own angles, and (**B**) Equation (18), where the rotor turning operator cycles the populations through the rotor orientations at every point on a two-angle spherical grid. Both approaches may be shown to be algebraically equivalent to, and about as fast computationally as, Floquet theory [5,18], with the notable difference that perturbative corrections to the rotating frame approximations are easier to implement within the Fokker-Planck formalism – Spinach [12] permits arbitrary-order corrections.*

An important feature of the operators in square brackets in Equations (16) and (18) is that they are time-independent and work as a part of the background Hamiltonian – *in the Fokker-Planck picture, magic angle spinning effectively looks like a static interaction.* All rotor cycle time slicing concerns intrinsic to the Liouville - von Neumann approach to MAS NMR simulation simply disappear. The practical convenience of this fact is hard to overstate – a good real-life example is given in our recent paper [46].

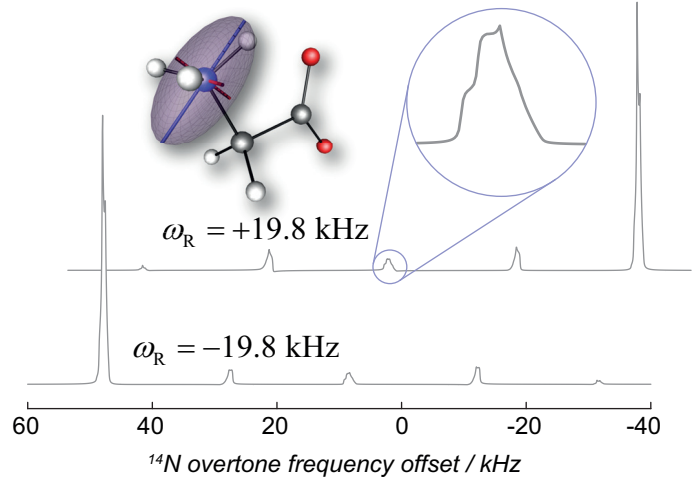

*Figure 2. An illustration of the dependence of the $^{14}$N quadrupolar overtone MAS NMR spectrum on the rotor spinning direction, simulated using Equation (18) with $C_Q$ = 1.18 MHz, $\eta_Q$ = 0.53 and a chemical shift of 32.4 ppm for the $^{14}$N nucleus, assuming proton decoupling. Full simulation details are given in our recent paper [46]. The ellipsoid plot illustrates the orientation of the $^{14}$N NQI tensor relative to the molecular geometry in the glycine zwitterion.*

Equation (16) has a further advantage – if the entire three-angle manifold is considered and the spatial basis set is chosen (following Freed's treatment of the mathematically related rotational diffusion process [8]) to be Wigner *D*-functions of those angles, Equation (16) *does not require a powder averaging grid*: the spherical average is obtained by taking the coefficient in front of $\mathfrak{D}_{0,0}^{(0)}$. Given the titanic amount of work that has gone into optimising spherical grids in NMR [47-54], this is remarkable. Further details on this "grid-free" formalism are given in our recent paper [5] – because matrix dimensions go up significantly, it may be viewed as striking a different balance between memory utilisation and CPU time.

At the numerical implementation level, the evolution generator is structured as a sparse block-diagonal matrix with each block corresponding to a point on the rotor phase grid. The job of the rotor turning generator is to move block populations around during evolution as illustrated in Figure 1B. The matrix form of Equation (18) at each point of the spherical grid is:

$$\frac{\partial}{\partial t}\begin{pmatrix}\boldsymbol{\rho}(\varphi_1,t)\\\boldsymbol{\rho}(\varphi_2,t)\\\vdots\\\boldsymbol{\rho}(\varphi_N,t)\end{pmatrix}=-i\begin{pmatrix}\mathbf{L}(\varphi_1)&&&\\&\mathbf{L}(\varphi_2)&&\\&&\ldots&\\&&&\mathbf{L}(\varphi_N)\end{pmatrix}\begin{pmatrix}\boldsymbol{\rho}(\varphi_1,t)\\\boldsymbol{\rho}(\varphi_2,t)\\\vdots\\\boldsymbol{\rho}(\varphi_N,t)\end{pmatrix}+\omega_\mathrm{R}\left(\left[\frac{\partial}{\partial\varphi}\right]\otimes\mathbf{1}\right)\begin{pmatrix}\boldsymbol{\rho}(\varphi_1,t)\\\boldsymbol{\rho}(\varphi_2,t)\\\vdots\\\boldsymbol{\rho}(\varphi_N,t)\end{pmatrix}\quad(21)$$

where $N$ is the number of points in the rotor phase grid, the expression for the rotor turning generator matrix $[\partial/\partial\varphi]$ is given in Equation (13) and $\mathbf{1}$ is a unit matrix of the same dimension as the spin Hamiltonian superoperator. External time-dependent events, such as shaped pulses, should be accounted for by adding the corresponding Hamiltonian to each block of the Liouvillian. Any rotating frame transformations or other Hamiltonian modifications should likewise be applied to each block. An extensively annotated numerical implementation of Equation (21) is available in the `singlerot.m` context function of *Spinach* [12], which performs all required procedures automatically.

It could be argued that a time-independent rotor turning generator is also a feature of Floquet theory [45]. The two approaches do not actually compete – Floquet theory is obtained from Equation (18) when the basis set is chosen to be complex exponentials of the rotor phase (a formal proof is given in [5]), meaning that Floquet theory is a special case of the Fokker-Planck formalism. The advantage of the finite grid approach is that Equation (21) does not become any more complicated when second- and higher-order corrections to the rotating frame transformation are required – the corrections simply need to be applied to every $\mathbf{L}(\varphi_j)$ block individually.

A good applications example is given by quadrupolar overtone simulations, where high-order perturbative corrections to rotating frame transformations are inevitable and a rather complicated effective Hamiltonian treatment must be performed to make it possible to simulate the spectrum in reasonable time [46]. Figure 2 illustrates the non-trivial dependence of the quadrupolar overtone MAS NMR spectrum on the spinning *direction* of the MAS rotor [55,56] – the sideband intensity pattern is reflected when the spinning direction is changed from clockwise to counter-clockwise. Further simulation details are given in [46], the extensively annotated Matlab code is a part of the standard example set supplied with *Spinach* [12].

## 4. Case study 2: multiple angle spinning NMR

The equation of motion for spin dynamics in a multiple angle spinning NMR experiment [57,58] is a simple extension of Equation (18). In the case of double rotation, we have two rotor turning generators:

$$\frac{\partial}{\partial t}\boldsymbol{\rho}(\varphi_O,\varphi_I,t) = \left[-i\mathbf{L}(\mathbf{n}_O,\mathbf{n}_I,\varphi_O,\varphi_I) + \omega_O\frac{\partial}{\partial \varphi_O} + \omega_I\frac{\partial}{\partial \varphi_I}\right]\boldsymbol{\rho}(\varphi_O,\varphi_I,t) \quad (22)$$

where $\mathbf{n}_O$ is the direction vector specifying the orientation of the outer rotor relative to the laboratory frame of reference, $\mathbf{n}_I$ is the direction vector specifying the orientation of the inner rotor relative to the outer rotor, $\varphi_{O,I}$ are outer and inner rotor phases, $\omega_{O,I}$ are outer and inner rotor angular frequencies, $\boldsymbol{\rho}$ is the state vector, and $\mathbf{L}$ is the spin dynamics Liouvillian comprising the superoperators for free evolution, relaxation theory and chemical kinetics:

$$\mathbf{L}(\mathbf{n}_O,\mathbf{n}_I,\varphi_O,\varphi_I,t) = \mathbf{H}(\mathbf{n}_O,\mathbf{n}_I,\varphi_O,\varphi_I) + i\mathbf{R} + i\mathbf{K} \quad (23)$$

The spin Hamiltonian commutation superoperator is:

$$\mathbf{H}\left(\mathbf{n}_O,\mathbf{n}_I,\varphi_O^{(n)},\varphi_I^{(k)},\alpha,\beta,\gamma\right) =$$
$$= \sum_{km}\left[\mathfrak{D}_{\text{out2lab}}(\mathbf{n}_O)\mathfrak{D}_{\text{rotor}}\left(\varphi_O^{(n)}\right)\mathfrak{D}_{\text{inn2out}}(\mathbf{n}_I)\mathfrak{D}_{\text{rotor}}\left(\varphi_I^{(k)}\right)\mathfrak{D}_{\text{crystal}}(\alpha,\beta,\gamma)\right]_{km}\mathbf{Q}_{km} \quad (24)$$

where the 25 irreducible spherical components $\mathbf{Q}_{km}$ of the Hamiltonian [44] are the same as those in Equation (20). The five sets of Wigner *D*-matrices in Equation (24) accomplish the following transformations. $\mathfrak{D}_{\text{crystal}}(\alpha,\beta,\gamma)$ rotates the molecule from the reference frame, in which the interactions are defined, into each of the individual crystallite orientations specified by the spherical averaging grid. $\mathfrak{D}_{\text{rotor}}\left(\varphi_I^{(k)}\right)$ then rotates the result, around the Z axis, into the orientation that corresponds to the specified inner rotor phase. $\mathfrak{D}_{\text{inn2out}}(\mathbf{n}_I)$ then rotates the result into the orientation that corresponds to the inner rotor direction vector $\mathbf{n}_I$ relative to the outer rotor. $\mathfrak{D}_{\text{rotor}}\left(\varphi_O^{(n)}\right)$ then rotates the result, around the Z axis, into the orientation that corresponds to the specified outer rotor phase. Finally, $\mathfrak{D}_{\text{out2lab}}(\mathbf{n}_O)$ rotates the result into the orientation that corresponds to the outer rotor direction vector $\mathbf{n}_O$ relative to the laboratory frame of reference. Matrix representations for the rotor turning generators $\partial/\partial \varphi_I$ and $\partial/\partial \varphi_O$ are given by Equation (13). Here it is also possible to perform inexpensive $\gamma$ angle and relative rotor phase averaging efficiently by averaging over the phase grid points of both rotors, meaning that only a two-angle spherical grid is required in many cases. Setting Equation (24) up numerically is a significant undertaking and the reader is therefore advised to inspect and use the extensively annotated `double-rot.m` context function of *Spinach* [12] that performs this procedure automatically.

The matrix form of Equation (22) looks similar to Equation (21), but with two rotor turning operators:

$$\frac{\partial}{\partial t}\begin{pmatrix} \rho(\varphi_O^{(1)},\varphi_I^{(1)},t) \\ \rho(\varphi_O^{(1)},\varphi_I^{(2)},t) \\ \vdots \end{pmatrix} = -i\begin{bmatrix} \begin{pmatrix} L(\varphi_O^{(1)},\varphi_I^{(1)}) & & \\ & L(\varphi_O^{(1)},\varphi_I^{(2)}) & \\ & & \ldots \end{pmatrix} + \omega_O \mathbf{D}_O + \omega_I \mathbf{D}_I \end{bmatrix}\begin{pmatrix} \rho(\varphi_O^{(1)},\varphi_I^{(1)},t) \\ \rho(\varphi_O^{(1)},\varphi_I^{(2)},t) \\ \vdots \end{pmatrix} \quad (25)$$

$$\mathbf{D}_O = [\partial/\partial\varphi]_{N_O} \otimes \mathbf{1}_{N_I} \otimes \mathbf{1}_{N_S}, \quad \mathbf{D}_I = \mathbf{1}_{N_O} \otimes [\partial/\partial\varphi]_{N_I} \otimes \mathbf{1}_{N_S}$$

where $N_O$ is the number of phase grid points for the outer rotor, $N_I$ is the number of phase grid points for the inner rotor, $N_S$ is the dimension of the spin Hamiltonian commutation superoperator, $\mathbf{1}$ are identity superoperators of indicated dimensions, $[\partial/\partial\varphi]$ are Fourier spectral differentiation matrices of indicated dimensions obtained using Equation (13), state vectors $\rho(\varphi_O^{(n)},\varphi_I^{(k)},t)$ are stacked vertically in the order of increasing index of the inner rotor phase grid points, followed by the increasing index of the outer rotor phase grid points, and the block-diagonal matrix in the middle is obtained by concatenating individual phase grid point Liouvillians $\mathbf{L}(\varphi_O^{(n)},\varphi_I^{(k)})$ in the same order as the state vectors. External time-dependent events, such as shaped pulses, should be accounted for by adding the corresponding Hamiltonian superoperator to each matrix block. Any rotating frame transformations or other Hamiltonian modifications should likewise simply be applied to each block.

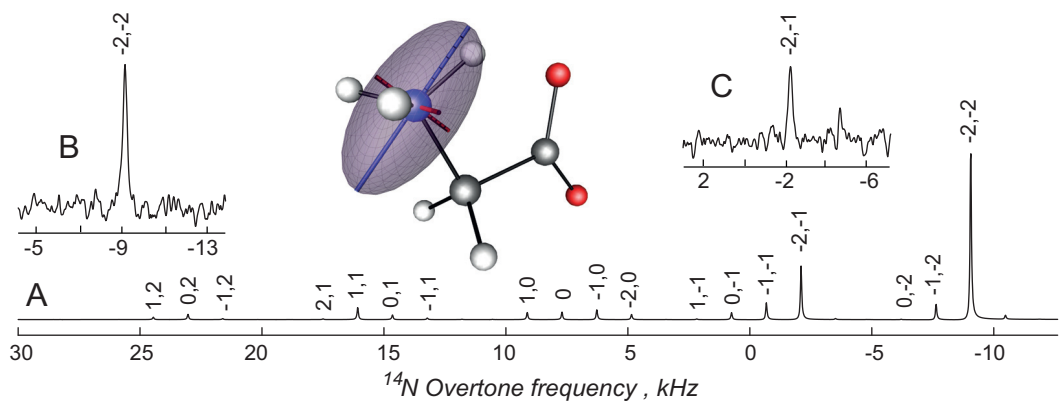

*Figure 3.* Double rotation (1425 Hz, 6950 Hz) $^{14}$N overtone NMR spectrum of glycine, reproduced from the paper by Carravetta and co-authors [59]. The simulation (**A**) was carried out using ideal pulses, but due to the low transition moment across the forbidden overtone transition, it is instrumentally impossible to excite more than one line at a time. Experimental overtone spectra (**B,C**) were therefore recorded selectively at the most intense spinning sidebands. The ellipsoid plot illustrates the orientation of the $^{14}$N NQI tensor relative to the molecular geometry in the glycine zwitterion.

Much as in the case of MAS described in the previous section, the increase in the matrix dimension is compensated for by the convenience of having a time-independent rotor turning generator – the $\omega_O \mathbf{D}_O + \omega_I \mathbf{D}_I$ term in Equation (25) is just another constant matrix. From the programming point of view, setting up a DOR pulse sequence is at this point no harder than setting it up for liquid state NMR. The fact that a complicated spinning process is present no longer generates any housekeeping difficulties because it is a part of what looks like a static background Hamiltonian.

A sophisticated DOR simulation example from our recent paper [59] is shown in Figure 3. Quadrupolar overtone simulations are slow [55,56], and the most efficient way of accelerating them (Equations 6 and 7 in [46]) requires the evolution generator to be time-independent, which it would not be in the Liouville - von Neumann formalism. However, Equation (22) fits that requirement perfectly.

## 5. Case study 3: spatio-temporal NMR experiments

A number of advanced magnetic resonance experiments (Thrippleton-Keeler zero-quantum filter [60], ultrafast NMR sequences [61], pure shift NMR sequences [62], slice selection in MRI [63], *etc.*) use frequency-modulated pulses in the presence of magnetic field gradients. A significant source of magnetization loss in such experiments is the inevitable presence of diffusion and hydrodynamics that reduce the efficiency of gradient refocusing steps. Good theoretical models of this process must simultaneously account for coordinate-dependent quantum mechanical spin evolution, diffusion and hydrodynamics. That is not a simple task within the Liouville - von Neumann formalism, where formally correct but computationally expensive Monte-Carlo sums over stochastic trajectories have so far been used [64].

In the simplest case, the Fokker-Planck equation for spin evolution under a radiofrequency pulse in the presence of a magnetic field gradient, translational diffusion and uniform flow in one dimension would borrow spatial dynamics generators from Equations (8) and (10):

$$\frac{\partial}{\partial t}\boldsymbol{\rho}(\varphi,z,t) = \left[-i\mathbf{L}(\varphi,z,t) + \omega(t)\frac{\partial}{\partial \varphi} + D_\mathrm{T}\frac{\partial^2}{\partial z^2} + v\frac{\partial}{\partial z}\right]\boldsymbol{\rho}(\varphi,z,t) \qquad (26)$$

where $D_\mathrm{T}$ is the translational diffusion coefficient, $v$ is the flow velocity, $z$ is the sample coordinate, $\varphi$ is the phase of the radiofrequency pulse and $\mathbf{L}(\varphi,z,t)$ is the spin Liouvillian

$$\begin{aligned}\mathbf{L}(\varphi,z,t) &= \mathbf{H}_0 + a(t)\left[\mathbf{S}_\mathrm{X}\cos(\varphi+\varphi_0) + \mathbf{S}_\mathrm{Y}\sin(\varphi+\varphi_0)\right] + \\ &\quad + g_\mathrm{Z}(t)z\sum_n \gamma_n \mathbf{S}_\mathrm{Z}^{(n)} + i\mathbf{K} + i\mathbf{R}\end{aligned} \qquad (27)$$

that contains the time-independent Hamiltonian $\mathbf{H}_0$ (centre point chemical shifts, *J*-couplings and other static interactions), the radiofrequency term with initial phase $\varphi_0$ and a time-dependent amplitude $a(t)$, a gradient term with a time-dependent amplitude $g_\mathrm{Z}(t)$, the chemical kinetics superoperator $\mathbf{K}$ and the relaxation superoperator $\mathbf{R}$. Here, the advantage of Equation (26) over the equivalent Liouville - von Neumann equation formulation is not in the elimination of time dependence, but in the fact that *diffusion and flow are constant operators working in the background*.

For numerical calculations using Equation (26) to become possible, matrix representations must be obtained for all of its constituent operators. Spin operators are matrices by definition. The differential operators acquire matrix representations when a grid of points is chosen for $z$ and $\varphi$. Finite difference matrices [32] are sufficient for the spatial coordinate and Equation (13) is recommended for the radiofrequency phase derivative operator. The resulting matrices must be projected into the Fokker-Planck space by taking Kronecker products with unit matrices on all coordinates that are unaffected by the operator. The procedure results in the following matrix representation for Equation (26):

$$\frac{\partial}{\partial t}\boldsymbol{\rho}(t) = \left[-i\mathbf{L}(t) + \omega(t)\mathbf{D}_\varphi + D_\mathrm{T}\mathbf{D}_z^2 + v\mathbf{D}_z\right]\boldsymbol{\rho}(t)$$

$$\mathbf{D}_\varphi = \left[\frac{\partial}{\partial \varphi}\right]_M \otimes \mathbf{1}_N \otimes \mathbf{1}_K \qquad \mathbf{D}_z = \mathbf{1}_M \otimes \left[\frac{\partial}{\partial z}\right]_N \otimes \mathbf{1}_K$$

(28)

in which $M$ is the number of points in the RF phase grid, $N$ is the number of points in the coordinate grid and $K$ is the dimension of the spin state space. The block structure of the spin Liouvillian is:

$$\mathbf{L}(t) = \begin{pmatrix} \mathbf{L}_{\varphi_1,z_1}(t) & & & \\ & \mathbf{L}_{\varphi_1,z_2}(t) & & \\ & & \ldots & \\ & & & \mathbf{L}_{\varphi_M,z_N}(t) \end{pmatrix}$$

$$\mathbf{L}_{\varphi_n,z_k}(t) = \mathbf{H}_0 + a(t)\left[\mathbf{S}_\mathrm{X}\cos(\varphi_n + \varphi_0) + \mathbf{S}_\mathrm{Y}\sin(\varphi_n + \varphi_0)\right] + \\ + g_\mathrm{Z}(t)z_k\sum_n \gamma_n \mathbf{S}_\mathrm{Z}^{(n)} + i\mathbf{K} + i\mathbf{R}$$

(29)

The responsibility for using sufficient number of grid points along both coordinates rests with the user – in practice the number of grid points is increased until the answer stops changing. At the end of the calculation, a partial trace over the RF phase coordinate and the sample position coordinate (*i.e.* a sum of all $(\varphi_k, z_m)$ blocks in $\boldsymbol{\rho}$) produces the average spin state that is visible to spectrometer coils.

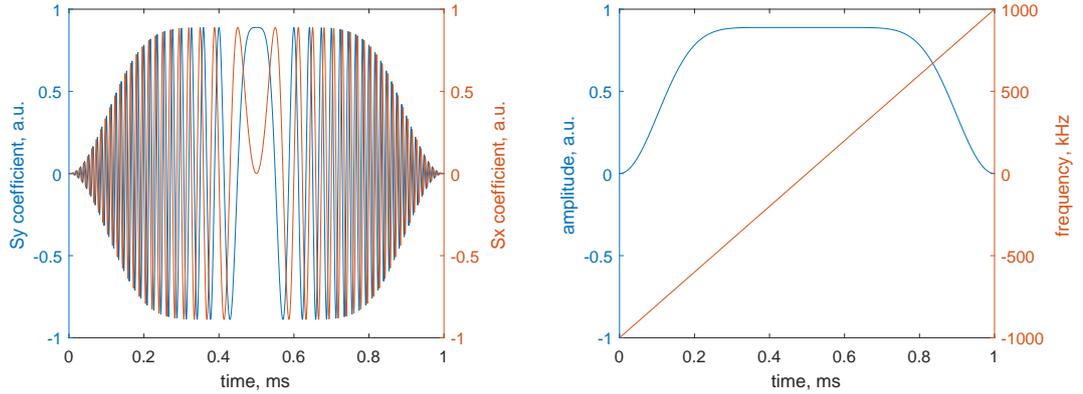

*Figure 4. Cartesian (left) vs. frequency-amplitude (right) representation of a frequency-swept pulse with the frequency range of 2 MHz around the centre and a duration of 1 ms.*

Equation (26) has the following advantages over other equivalent formulations:

1. The radiofrequency pulse appears in frequency-amplitude, rather than Cartesian, parameterisation. The difference between the two is illustrated in Figure 4. It is clear that frequency-swept pulses are easier to discretise in frequency-amplitude coordinates.

2. Diffusion and flow generators are constant matrices that may simply be viewed as a part of the background Hamiltonian. For all practical purposes, the calculation looks like a standard simulation of a shaped radiofrequency pulse in NMR spectroscopy. Expensive Monte-Carlo averages over stochastic trajectories are unnecessary.

3. Diffusion and flow are non-periodic motions – while it could be argued in the sections above that Floquet theory provides an established alternative for periodic processes in the spin Hamiltonian, that is not the case here.

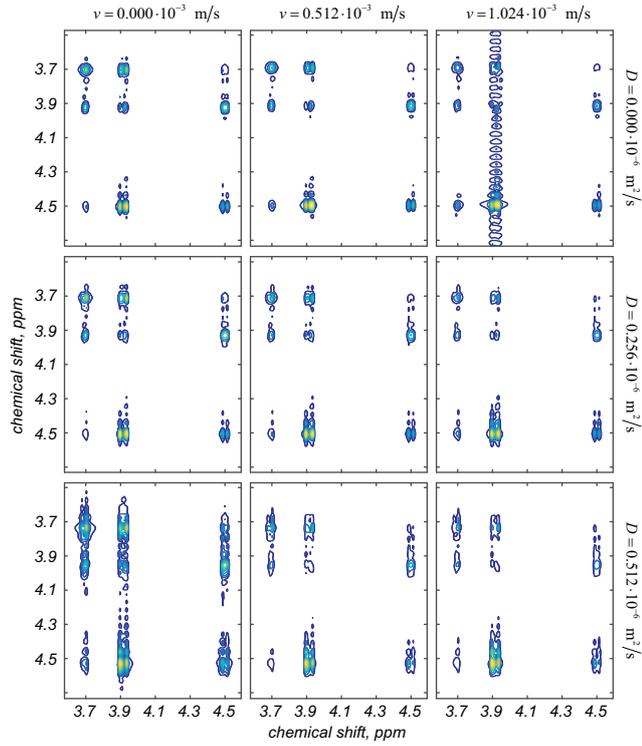

*Figure 5. Ultrafast COSY [61] spectrum of a three-spin system ($\delta_1$ = 3.70 ppm, $\delta_2$ = 3.92 ppm, $\delta_1$ = 4.50 ppm, $J_{1,2}$ = 10 Hz, $J_{2,3}$ = 12 Hz, $J_{1,3}$ = 4 Hz) as a function of the diffusion coefficient and the linear flow velocity. The spectrum was simulated for a sample distributed in one spatial dimension with a 500-point grid with absorptive boundary conditions used to discretise a 15 mm interval representing the active volume of a typical NMR sample. The ultrafast COSY sequence was recorded using 128 gradient readout loops, 512 acquisition points each, with a dwell time of 1 µs. Acquisition gradient amplitude was set to 0.10 T/m, encoding gradient amplitude to 0.01 T/m, and the coherence selection gradient amplitude to 0.47 T/m. The duration of the coherence selection gradient pulses was 1 ms. Spatial encoding was achieved with a WURST pulse with a smoothing factor of 40 and a bandwidth of 10 kHz acting over a 15 ms time interval and discretised over a 1000-point time grid. Technical details about ultrafast COSY simulations in Spinach are available in [65].*

A practical example of Equation (26) enabling the simulation of a sophisticated spatio-temporal NMR experiment is ultrafast NMR spectroscopy [61,65]. The processes that make ultrafast NMR possible take place in the direct product $\mathbb{F}$ of the one-dimensional real space $\mathbb{R}$ and the spin state space $\mathbb{S}$:

$$\mathbb{F} = \mathbb{R} \otimes \mathbb{S} \tag{30}$$

Once the spatial grid is chosen, the Fokker-Planck equation of motion for an ensemble of spin systems diffusing and flowing in one spatial dimension in the presence of a magnetic field gradient acquires the following matrix form:

$$\frac{d}{dt}\boldsymbol{\rho}(t) = -i\mathbf{F}(t)\boldsymbol{\rho}(t)$$
$$\mathbf{F}(t) = \mathbf{1}_\mathbb{R} \otimes \mathbf{H}(t) + g_Z(t)\mathbf{Z} \otimes \sum_n \gamma_n \mathbf{S}_Z^{(n)} + \tag{31}$$
$$+ iD_T \mathbf{D}_Z^2 \otimes \mathbf{1}_\mathbb{S} + iv\mathbf{D}_Z \otimes \mathbf{1}_\mathbb{S} + i\mathbf{1}_\mathbb{R} \otimes \mathbf{R} + i\mathbf{1}_\mathbb{R} \otimes \mathbf{K}$$

where $\mathbf{H}$ is the spin Hamiltonian commutation superoperator, $g_Z(t)$ is the gradient amplitude, $\mathbf{Z}$ is the spatial coordinate operator (a diagonal matrix with the coordinates of grid points on the diagonal), $\gamma_k$ are magnetogyric ratios of the spins in the system, $\mathbf{S}_Z^{(k)}$ are the corresponding longitudinal spin commutation superoperators, $D_T$ is the diffusion coefficient, $v$ is the flow velocity, $\mathbf{D}_Z$ is a matrix representation of the finite difference first derivative operator on our grid [32], $\mathbf{R}$ is the spin relaxation superoperator, $\mathbf{K}$ is the chemical kinetics superoperator, $\mathbf{1}_\mathbb{S}$ is a unit spin superoperator and $\mathbf{1}_\mathbb{R}$ is the unit operator on the spatial grid. This equation may be solved using standard time propagation methods [66].

Figure 5 demonstrates the effect of diffusion and flow on the ultrafast COSY spectrum of a simple three-spin system. The *J*-modulation effects – a quantum mechanical phenomenon that essentially requires full density matrix treatment in the spin subspace – are clearly visible [65]. Another interesting conclusion is that fast diffusion and fast flow can cancel each other's deleterious effects to some extent. This is likely because they compensate each other for some small fraction of the sample.

A three-dimensional version of Equation (26) would include anisotropic diffusion and hydrodynamics [67]. The diffusion term is just $\nabla^T \cdot \mathbf{D}_T \cdot \nabla$, where $\mathbf{D}_T$ is the translational diffusion tensor. However, hydrodynamics equations are in general non-linear and therefore undesirable within the linear structure of Equation (26). Their solution is a difficult task that requires specialised numerical methods [68]; the prospect of having to integrate a Navier-Stokes solver into a magnetic resonance package is dire. Fortunately, an elegant workaround is available because flows in magnetic resonance systems are usually stationary [69,70]. A hydrodynamics solver may be run to obtain the stationary velocity field $\mathbf{v}(\mathbf{r})$ which may then be combined with Equation (5) to yield the following extra term in the evolution generator:

$$\nabla_\mathbf{r} \cdot \left[ \mathbf{v}(\mathbf{r}) \boldsymbol{\rho}(\mathbf{r},t) \right] = \left( \nabla_\mathbf{r} \cdot \mathbf{v}(\mathbf{r}) \right) \boldsymbol{\rho}(\mathbf{r},t) + \mathbf{v}(\mathbf{r}) \cdot \nabla_\mathbf{r} \boldsymbol{\rho}(\mathbf{r},t) = \\ = \left[ \nabla_\mathbf{r} \cdot \mathbf{v}(\mathbf{r}) + \mathbf{v}(\mathbf{r}) \cdot \nabla_\mathbf{r} \right] \boldsymbol{\rho}(\mathbf{r},t) \quad (32)$$

The operator in square brackets does not depend on time and therefore fits naturally into the background evolution operator of Equation (26). Its matrix representation is easy to obtain using finite difference matrices on a suitable grid. In the case of one-dimensional flow of constant velocity, the flow term from Equation (26) is produced.

## 6. Case study 4: orientation selection in double-electron resonance

Another relevant example where the Fokker-Planck approach is advantageous is the off-resonance irradiation and orientation selection that take place during DEER experiments [71]. Because large zero-field splittings and soft microwave pulses at two distinct frequencies may be involved [72], Liouville - von Neumann formalism simulations must proceed by explicitly discretising both waveforms and running a torturously slow simulation with a time-dependent Hamiltonian in the laboratory frame. However, the Fokker-Planck evolution generator, coming from Equation (18), is again time-independent:

$$\frac{\partial}{\partial t} \boldsymbol{\rho}(\varphi, x, t) = \left[ -i\mathbf{L}(\varphi, x) + \omega_{MW} \frac{\partial}{\partial \varphi} \right] \boldsymbol{\rho}(\varphi, x, t) \quad (33)$$

where the Liouvillian now includes the background Hamiltonian and the microwave irradiation with the phase $\varphi$ that is incremented at the frequency $\omega_{MW}$:

$$\mathbf{L}(\varphi, x, t) = \mathbf{H}_0 + a \left[ \mathbf{S}_X \cos(\varphi + \varphi_0) + \mathbf{S}_Y \sin(\varphi + \varphi_0) \right] + i\mathbf{K} + i\mathbf{R} \quad (34)$$

where $\mathbf{H}_0$ is the spin Hamiltonian containing all interactions intrinsic to the spin system, $\{\mathbf{S}_X, \mathbf{S}_Y\}$ are electron spin superoperators, $a$ is the amplitude of the microwave pulse and $\varphi_0$ is its initial phase. A matrix representation of this problem is obtained by discretising $\partial/\partial\varphi$ using Equation (13):

$$\frac{\partial}{\partial t} \boldsymbol{\rho}_\varphi(t) = \left[ -i\mathbf{L}_\varphi + \omega_{MW} \mathbf{D}_\varphi \right] \boldsymbol{\rho}_\varphi(t), \qquad \mathbf{D}_\varphi = \left[ \frac{\partial}{\partial \varphi} \right]_M \otimes \mathbf{1}_K \quad (35)$$

in which $M$ is the number of points in the microwave phase grid (fewer than ten are required in practice) and $K$ is the dimension of the spin state space. The block structure of the spin Liouvillian is:

$$\mathbf{L}_\varphi = \begin{pmatrix} \mathbf{L}_{\varphi_1} & & & \\ & \mathbf{L}_{\varphi_2} & & \\ & & \ldots & \\ & & & \mathbf{L}_{\varphi_M} \end{pmatrix} \tag{36}$$

$$\mathbf{L}_{\varphi_n} = \mathbf{H}_0 + a\left[\mathbf{S}_X \cos(\varphi_n + \varphi_0) + \mathbf{S}_Y \sin(\varphi_n + \varphi_0)\right] + i\mathbf{K} + i\mathbf{R}$$

Equation (33) should be invoked for each of the three microwave pulses; the usual Liouville - von Neumann equation is sufficient for the free evolution period. A partial trace over the microwave phase grid at the end of each pulse produces the resulting spin density matrix.

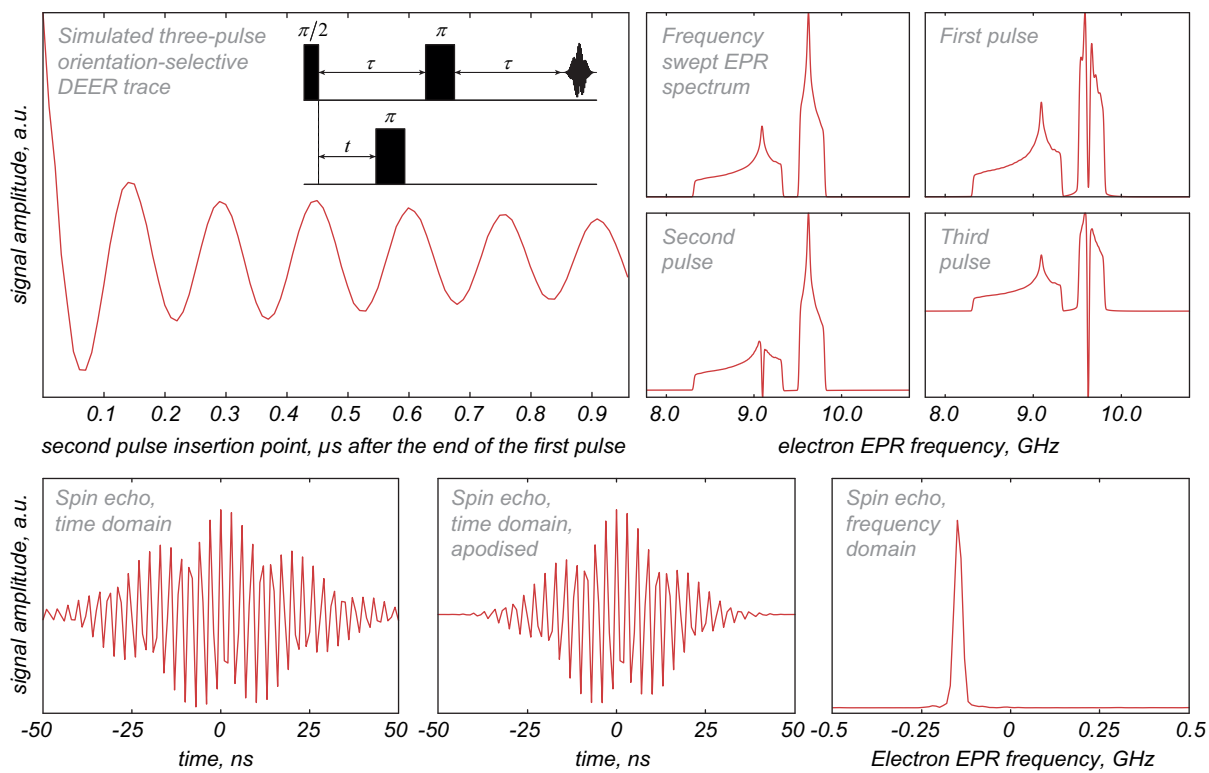

*Figure 6.* A three-pulse DEER trace and associated diagnostic information for a two-electron system, simulated as described in the main text, at a magnetic induction of 0.34518 Tesla with the following parameters: first electron g-tensor eigenvalues [2.284, 2.123, 2.075], Euler angles [45°, 90°, 135°], second electron g-tensor eigenvalues [2.035, 2.013, 1.975], Euler angles [120°, 60°, 30°], inter-electron distance of 20 Angstrom along the X axis, pulse durations 20 ns, 40 ns and 40 ns, pulse amplitude 8 MHz for all three pulses, pulse frequencies 9.623 GHz, 9.092 GHz and 9.623 GHz, 1 µs gap between the first and the third pulse, 100 steps in the position of the second pulse, 100 points in the echo sampling window. The DEER trace has a modulation depth of approximately 10% of the total echo intensity. The associated Matlab source code is available in the example set supplied with versions 1.8 and later of Spinach library [12].

Equation (33) is currently the default formalism for soft-pulse DEER simulation in *Spinach* [12] – the corresponding functions are extensively annotated and should be consulted for the details of the numerical implementation. One of the examples included with the package is detailed in Figure 6. It is clear that the description of the soft pulses is very accurate – even the *sinc* wiggles are visible.

## 7. Case study 5: overtone cross-polarisation under MAS

Another good example of the Fokker-Planck formalism dealing successfully with a combination of spatial motion, quantum mechanics and radiofrequency irradiation is quadrupolar overtone cross-polarisation. Even basic overtone NMR simulations are difficult because they involve a radiofrequency that cannot be assumed to be close to the Zeeman frequency of the nucleus in question [46,55,56,59]. This impedes the first step normally taken in magnetic resonance simulations – the rotating frame transformation – and makes Liouville - von Neumann equation simulations slow. By that standard, *cross-polarisation* overtone MAS simulations are the stuff of nightmares – there are three non-commensurate and very different frequencies (MAS, $^1$H, $^{14}$N overtone) in the time-dependent part of the Hamiltonian. The system must be propagated for several milliseconds in the laboratory frame and the entire calculation averaged over a three-angle spherical grid. The situation may be simplified somewhat by going into the partial rotating frame with respect to the proton Zeeman frequency and taking care to retain the pseudosecular terms in the various couplings involving protons, but the two remaining frequencies are still problematic.

The advantage of the Fokker-Planck formulation in this context is threefold. Firstly, it makes time dependence disappear completely from the evolution generator because phases of the RF and the rotor are now incremented by time-independent derivative terms introduced in Equations (26) and (18). Secondly, only a two-angle spherical grid is required because rotor phase averaging is built-in. Thirdly, the lack of explicit time dependence in the evolution generator permits frequency domain detection which, for overtone peaks, is much faster than the time domain approach [46,59]. The equation of motion is:

$$\frac{\partial}{\partial t}\mathbf{\rho}(\Omega,t) = \left[ -i\mathbf{L}(\Omega) + \omega_{\text{MAS}} \frac{\partial}{\partial \varphi_{\text{MAS}}} + \omega_{\text{RF}}^{\text{N}} \frac{\partial}{\partial \varphi_{\text{RF}}^{\text{N}}} \right] \mathbf{\rho}(\Omega,t) \qquad (37)$$

with the following spin Liouvillian (assuming, to match the available experimental literature, $^{14}$N to be the quadrupolar nucleus and $^1$H to be the polarisation source):

$$\mathbf{L}(\Omega) = \mathbf{H}_0 + \sum_{k,n,m=-2}^{2} \mathfrak{D}_{kn}^{(2)}(\mathbf{r}_{\text{MAS}},\varphi_{\text{MAS}}) \mathfrak{D}_{nm}^{(2)}(\alpha,\beta,\gamma) \mathbf{Q}_{km} +$$
$$+ a_{\text{RF}}^{\text{N}} \left[ \mathbf{S}_{\text{Z}}^{\text{N}} \cos\theta + \left( \mathbf{S}_{\text{X}}^{\text{N}} \cos(\varphi_{\text{RF}}^{\text{N}}) + \mathbf{S}_{\text{Y}}^{\text{N}} \sin(\varphi_{\text{RF}}^{\text{N}}) \right) \sin\theta \right] \qquad (38)$$
$$+ a_{\text{RF}}^{\text{H}} \left[ \mathbf{S}_{\text{Z}}^{\text{H}} \cos\theta + \mathbf{S}_{\text{X}}^{\text{H}} \sin\theta \right] + i\mathbf{R} + i\mathbf{K}$$

where $\Omega = \{\alpha,\beta,\gamma,\varphi_{\text{MAS}},\varphi_{\text{RF}}^{\text{N}}\}$, $\mathbf{\rho}$ is the density matrix, $\theta$ is the magic angle, $\varphi_{\text{MAS}}$ is the MAS rotor phase, $\varphi_{\text{RF}}^{\text{N}}$ is the nitrogen radiofrequency phase, $\omega_{\text{MAS}}$ and $\omega_{\text{RF}}^{\text{N}}$ are the corresponding frequencies, $\mathbf{H}_0$ is the orientation-independent part of the Hamiltonian commutation superoperator (*J*-couplings, isotropic chemical shifts, *etc.*), $\mathbf{Q}_{km}$ are the 25 irreducible spherical components [44] of the anisotropic part (dipolar interactions, quadrupolar interactions, chemical shift anisotropies, *etc.*), $\mathfrak{D}_{nm}^{(2)}(\alpha,\beta,\gamma)$ are second rank Wigner *D*-functions of the three Euler angles specifying the crystallite orientation within the powder grid [73], $\mathfrak{D}_{kn}^{(2)}(\mathbf{r}_{\text{MAS}},\varphi_{\text{MAS}})$ are second rank Wigner *D*-functions defining rotor orientation and phase using angle-axis parameterisation [74] with $\vec{r}_{\text{MAS}} = \begin{bmatrix} \sqrt{2/3} & 0 & \sqrt{1/3} \end{bmatrix}$, $\{\mathbf{S}_{\text{X}}^{\text{H}},\mathbf{S}_{\text{Y}}^{\text{H}},\mathbf{S}_{\text{Z}}^{\text{H}}\}$ are proton Zeeman commutation superoperators, $\{\mathbf{S}_{\text{X}}^{\text{N}},\mathbf{S}_{\text{Y}}^{\text{N}},\mathbf{S}_{\text{Z}}^{\text{N}}\}$ are nitrogen Zeeman commutation superoperators, $a_{\text{RF}}^{\text{H}}$ is proton RF amplitude and $a_{\text{RF}}^{\text{N}}$ is nitrogen RF amplitude.

The matrix representation for Equation (37) is built in a similar way to the previous sections, by discretising the RF phase and the rotor phase on finite grids and using Equation (13) to obtain matrix representations for the phase derivative operators. In practical calculations on $^1$H-$^{14}$N two-spin systems in glycine and N-acetylvaline (the experimental paper will be published in due course), 15 points were found to be sufficient for the rotor phase grid and 5 points for the RF phase grid, meaning that the total spatial dynamics subspace dimension is 75. Given the Liouville space dimension of 36 for the spin subspace, the total dimension for the matrix representation of Equation (37) is 2700, which is easy to handle using modern sparse matrix manipulation methods, and likely tensor train methods in due course [75,76].

Because the time evolution generator in Equation (37) is time-independent, time propagation for the duration of the cross-polarisation period may be carried out with a single matrix exponential:

$$\boldsymbol{\rho}(t_{CP}) = \exp[-i\mathbf{F}t_{CP}]\boldsymbol{\rho}(0) \tag{39}$$

where $\mathbf{F}$ is the matrix representation of the Fokker-Planck evolution generator. Because the complexity of the scaling and squaring procedure [77,78] for matrix exponentiation is asymptotically logarithmic in the duration of the time interval $t_{CP}$, the fact that $t_{CP} \gg \|\mathbf{H}\|^{-1}$ is not in practice a problem [43].

Once the cross-polarisation pulse is over, the equation of motion becomes an instance of Equation (18):

$$\frac{\partial}{\partial t}\boldsymbol{\rho}(\Omega,t) = \left[-i\mathbf{L}(\Omega) + \omega_{MAS}\frac{\partial}{\partial \varphi_{MAS}}\right]\boldsymbol{\rho}(\Omega,t) \tag{40}$$

with a Liouvillian that no longer contains any radiofrequency terms:

$$\mathbf{L}(\Omega) = \mathbf{H}_0 + \sum_{k,n,m=-2}^{2} \mathfrak{D}_{kn}^{(2)}(\mathbf{r}_{MAS},\varphi_{MAS})\mathfrak{D}_{nm}^{(2)}(\alpha,\beta,\gamma)\mathbf{Q}_{km} + i\mathbf{R} + i\mathbf{K} \tag{41}$$

where now $\Omega = \{\alpha,\beta,\gamma,\varphi_{MAS}\}$. For each point on the spherical averaging grid, the matrix representation of this equation is identical to Equation (21). If we denote the Fokker-Planck evolution generator $\mathbf{F}$, the following equation gives the Fourier transform of the free induction decay on $^{14}$N:

$$f(\omega) = \int_0^\infty \langle \mathbf{S}_+^N | \exp(-i\mathbf{F}t) | \boldsymbol{\rho}_0 \rangle e^{-i\omega t} dt = \langle \mathbf{S}_+^N | \left[\int_0^\infty \exp(-i\mathbf{F}t) e^{-i\omega t} dt\right] | \boldsymbol{\rho}_0 \rangle = \\ = -i \langle \mathbf{S}_+^N | (\mathbf{F} + \omega \mathbf{1})^{-1} | \boldsymbol{\rho}_0 \rangle \tag{42}$$

where $\mathbf{1}$ is a unit matrix of the same dimension as $\mathbf{F}$. The matrix dimension at this stage is smaller than at the cross-polarisation stage because a partial trace over the RF degrees of freedom is taken at the end of the CP period and the Liouvillian in Equation (42) no longer contains RF terms.

A significant advantage of Equation (42) over time-domain detection is that very few points are required (a few hundred in the frequency domain compared to billions in the time domain) and that the sparse matrix-inverse-times-vector operation is very fast when modern iterative solvers, such as ILU preconditioned GMRES [79], are used. Because a time-independent evolution generator is a requirement, this is only possible within the Fokker-Planck formalism.

## 8. Potential further applications

This section contains a speculative overview of other types of magnetic resonance simulations and applications that could be enabled, or made simpler, by the Fokker-Planck formalism. One such area is optimal control, where the problem of the interpretation of numerically optimised microwave and radiofrequency pulses is significant because their immediate appearance is obscure [44]. All current techniques rely either on the time-frequency representation of the pulse itself [80], or on the analysis of the spin system trajectory under the pulse [44]. In both cases the analysis step is performed *after* the optimal control solution has been obtained. The Fokker-Planck formalism is potentially useful here because it makes it possible to formulate the commonly used gradient ascent pulse engineering (GRAPE) algorithm [81,82] directly in the frequency-amplitude representation.

The standard setting for the quantum optimal control problem [81] involves a system with the Hamiltonian partitioned into the "drift" part $\mathbf{H}_0$ that we cannot influence, and the "control" part, in which the spectrometer can vary the coefficients $c_k(t)$ in front of some operators $\mathbf{H}_k$ [81]:

$$\mathbf{H}(t) = \mathbf{H}_0(t) + \sum_k c_k(t) \mathbf{H}_k \tag{43}$$

Analytical or numerical optimization of "control sequences" $c_k(t)$ to achieve some pre-defined experimental objectives is the subject of optimal control theory [83]. In most magnetic resonance cases, the drift Hamiltonian contains the magnet Zeeman terms and the spin-spin coupling terms. The control channels correspond to radiofrequency and microwave irradiation:

$$\mathbf{H}(t) = \underbrace{\sum_k \vec{\mathbf{S}}_k \cdot \mathbf{Z}_k \cdot \vec{B} + \frac{1}{2} \sum_{n,k} \vec{\mathbf{S}}_n \cdot \mathbf{A}_{nk} \cdot \vec{\mathbf{S}}_k}_{\text{drift}} + \underbrace{\sum_k \left[ c_k^{(X)}(t) \mathbf{S}_X^{(k)} + c_k^{(Y)}(t) \mathbf{S}_Y^{(k)} \right]}_{\text{X and Y controls}} \tag{44}$$

where the summation indices run over all spins in the system, $\mathbf{Z}_k$ are Zeeman tensors, $\mathbf{A}_{nk}$ are coupling tensors and $\vec{\mathbf{S}}_k = \{\mathbf{S}_X^{(k)}, \mathbf{S}_Y^{(k)}, \mathbf{S}_Z^{(k)}\}$ are spin operators. The current wisdom, illustrated well in [80], and also in Figure 4, is that Cartesian representation of the waveform $\{c_k^{(X)}(t), c_k^{(Y)}(t)\}$ is harder to interpret than amplitude-phase $\{a_k(t), \varphi_k(t)\}$ or amplitude-frequency $\{a_k(t), \omega_k(t)\}$ representation. In fact, this last one appears to be the most convenient [80].

It follows from Equation (26) that the amplitude-frequency representation appears naturally within the Fokker-Planck formalism:

$$\hat{H}(t) = \underbrace{\sum_k \vec{\mathbf{S}}_k \cdot \mathbf{Z}_k \cdot \vec{B} + \frac{1}{2} \sum_{n,k} \vec{\mathbf{S}}_n \cdot \mathbf{A}_{nk} \cdot \vec{\mathbf{S}}_k}_{\text{drift}} + \underbrace{\sum_k a_k(t) \left[ \mathbf{S}_X^{(k)} \cos(\varphi^{(k)}) + \mathbf{S}_Y^{(k)} \sin(\varphi^{(k)}) \right]}_{\text{amplitude controls}} + \underbrace{\sum_k \omega_k(t) \frac{\partial}{\partial \varphi^{(k)}}}_{\text{frequency controls}} \tag{45}$$

Pulse amplitudes and frequencies enter this equation linearly and may therefore be optimised by the GRAPE procedure [82], as well as its recent enhancements [81,84,85]. The first group of control operators consists of $\mathbf{S}_X^{(k)} \cos(\varphi^{(k)}) + \mathbf{S}_Y^{(k)} \sin(\varphi^{(k)})$ terms discretised on a finite phase grid:

$$\cos(\mathbf{\Phi}) \otimes \mathbf{S}_X^{(k)} + \sin(\mathbf{\Phi}) \otimes \mathbf{S}_Y^{(k)}, \qquad \mathbf{\Phi} = \begin{bmatrix} \varphi_1^{(k)} & 0 & \cdots & 0 \\ 0 & \varphi_2^{(k)} & \cdots & 0 \\ \vdots & \vdots & \ddots & 0 \\ 0 & 0 & 0 & \varphi_N^{(k)} \end{bmatrix} \tag{46}$$

The second group of control operators contains spectral finite difference representations of the derivative operators obtained using Equation (13). With those matrix representations in place, Equation (45) acquires the form that is directly usable by the existing optimal control software [12,17,86].

In our testing so far, this frequency-amplitude representation for control sequences did not significantly outperform either the Cartesian or the phase-amplitude control version of GRAPE. It should, however, certainly be mentioned here and kept on the books until such time as a system presents itself for which amplitude-frequency coordinates would be in some sense natural.

Another interesting, if exotic, hypothetical case of a spatial variable acting as a control channel in magnetic resonance experiments is the rotor phase. As per Equation (18), the spinning operator enters the Fokker-Planck equation of motion for a MAS experiment in the same algebraic way as the radiofrequency phase enters Equation (45). The matrix representation of the rotor phase derivative may therefore be used as a control operator and the associated spinning rate as a control coefficient. Another control coefficient is provided by the directional cosine of the spinning axis.

While variable spinning angle experiments are firmly established in NMR [87,88] and some applications of variable spinning rate experiments do exist [89], it is at this point unclear whether the use of optimal control theory here would result in any improvement, or whether the hardware of sufficient agility to control both the angle and the rate in real time could be built at a reasonable financial and time cost. We have not explored this matter any further, but it bears notice that the possibility exists.

## 8. Conclusions and outlook

Our experience of designing and coding *Spinach* [12] indicates that a high level of abstraction in the fundamental equations of motion, exotic and unwieldy though they may at first appear, is worth it in the long run because the resulting framework is general, flexible, extensible and maintainable. Throughout the last ten years we had to resist significant temptation to introduce case-specific tweaks that could have made some simulations faster at the cost of fragmenting the code into incompatible chunks. The result of this holistic approach is a software package that simulates everything there is in magnetic resonance. One factor that made this possible is *Matlab*. The other is the Fokker-Planck equation.

A formalism that simultaneously supports orientation-selective DEER [72], ultrafast NMR [61,65], singlet state flow imaging [64] and gadolinium cross-effect DNP (see the example set supplied with *Spinach* [12]), with an option of including the most general spin relaxation theory in existence [8], is very valuable. It constitutes a sweeping generalisation that also supports multi-site exchange problems [90], spin isomers [91], quantised spatial degrees of freedom [28], contains Floquet theory [45] and stochastic Liouville equation [8] as special cases, and accommodates the whole body of theoretical methods for NMR spectroscopy of diffusion and transport in complex media [69,70]. The basic principles have been known since the seminal papers and books by Fokker [22], Planck [23], Callaghan [69,70] and particularly Freed [8,24], but it was not until recently that the computers became powerful enough, matrix dimension reduction tools arrived [38-41] and sparse matrix libraries became sufficiently convenient. In our opinion, the Fokker-Planck equation should now be preferred to the Liouville - von Neumann formalism as the fundamental equation of motion for numerical simulations in magnetic resonance.

## Acknowledgements

This work was made possible by the funding from EPSRC (EP/M023664/1, EP/N006895/1). The author is grateful to Marina Carravetta, Maria Concistre, Jean-Nicolas Dumez, Jack Freed, David Goodwin, Olivier Lafon, Malcolm Levitt, Gareth Morris, Alex Nevzorov, Giuseppe Pileio and Phil Williamson for stimulating discussions, as well as the *Spinach* user community for the much-needed occasional nudge. The University of Southampton has slowed this work down by conducting a three-month disciplinary investigation against IK for working there during the Christmas break. The hospitality of the NMR group at the University of Lille, which hosted IK as a Visiting Professor over the Easter break, is warmly acknowledged.